\documentclass[reprint,
 amsmath,amssymb,
 aps,
prb,
floatfix,
]{revtex4-2}

\usepackage{graphicx}
\usepackage{dcolumn}
\usepackage{bm}

\usepackage{xcolor}
\usepackage{siunitx}

\begin{document}

\preprint{APS/123-QED}

\title{Localizing fractional quasiparticles on graphene quantum hall antidots}

\author{S.M. Mills}

\author{D.V. Averin}
\email{dmitri.averin@stonybrook.edu}

\author{X. Du}
\email{xu.du@stonybrook.edu}

\affiliation{Department of Physics, Stony Brook University, Stony Brook, New York 11794 USA}

\date{\today}

\begin{abstract}
We report localization of fractional quantum Hall (QH) quasiparticles on graphene antidots. By studying coherent tunneling through the localized QH edge modes on the antidot, we measured the QH quasiparticle charges to be approximately $\pm e/3$ at fractional fillings of $\nu = \pm 1/3$. The Dirac spectrum in graphene allows large energy scales and robust quasiparticle localization against thermal excitation. The capability of localizing fractional quasiparticles on QH antidots brings promising opportunities for realizing anyon braiding and novel quantum electronics. 
\end{abstract}

\maketitle

The fractional QH effect (FQHE) hosts quasiparticles with nontrivial anyonic exchange statistics \cite{PhysRevLett.53.722, PhysRevLett.52.1583}, inviting the possibility of novel quantum information devices based on the braiding of individual quasiparticles. Anyon braiding has been previously explored in QH interferometers \cite{PhysRevLett.98.076805, PhysRevB.72.075342, PhysRevLett.108.256804, braidingarXiv}, by engineering the pathways of the one-dimensional (1D) QH edge modes. The zero-dimensional (0D) QH antidot is another approach which, compared to QH interferometers, allow the possibility of control over individual quasiparticles. In a QH antidot, QH edge modes encircling a small void inside the 2-dimensional electron system (2DES) form quantized, discreet energy levels due to quantum confinement. Individual quasiparticles of the QH fluid can then be added to, extracted from, or transferred between QH antidots through tunneling \cite{Goldman1010, PhysRevLett.83.160}. The presence of discrete quantization energy levels minimizes the quasiparticle scattering phase space, optimizing coherence. Such approach therefore potentially allows manipulation of quasiparticles with non-trivial topological properties and  realization of novel quantum devices \cite{AVERIN200125}.\par
Prior works on fractionally charged quasiparticles largely focus on semiconductor-based 2DES in the FQH regime. The fractional charge of $\nu = 1/3$ quasiparticles has been studied in current partition noise of FQHE edge states \cite{PhysRevLett.79.2526, 1997Natur.389..162D}, and in DC transport measurements of GaAs-based QH antidots \cite{Goldman1010} and QH interferometers \cite{PhysRevLett.98.076805, PhysRevLett.108.256804}. More recently, charge transport at fractional fillings of  $\nu=1/3$ and $\nu=2/3$ has been investigated in a GaAs-based 2DES in the Aharonov-Bohm (AB) regime in a QH interferometer \cite{Nakamura2019}. Despite these developments, it is consistently found that the relevant energy scales, including Landau level (LL) spacing and the quantum confinement-induced energy level spacing, are small in GaAs-based 2DES. This necessitates challenging experimental conditions, including sub-100mK temperatures and ultra-low-noise electronics to realize transport in individual devices, and effectively precludes experiments with more complex multi-device structures even with the highest quality samples. \par

The limiting energy scale in a QH antidot is the quantization energy spacing: $\Delta \epsilon \sim \hbar v/D$. Here, the edge mode velocity $v$ is limited by the Fermi velocity and is proportional to the sharpness of the confinement potential, and $D$ is the diameter of the antidot. Given the size of the antidot is much larger than the magnetic length, a large edge mode velocity facilitates large quantization energy scales. Graphene as a 2D Dirac semimetal offers several advantages over GaAs including large QH gaps in both integer and fractional fillings \cite{RevModPhys.81.109, PhysRevLett.121.226801}, high precision in lithographic definition, and a large intrinsic Fermi velocity. The Dirac spectrum leads to different electric screening properties and Landau level structure from those of a conventional 2DEG, calling for investigations in QH interferometer and antidot setups. In recent years, mono-layer \cite{Weie1700600} and bi-layer graphene \cite{Allen2012, PhysRevLett.122.146801} interferometers have been used to study the dynamics of QH edge states, and large edge mode velocity has been measured in magnetoplasmons over long graphene edges \cite{PhysRevLett.110.016801}. More recently, we have demonstrated a graphene QH antidot which localizes $\nu = 2$ quasiparticles, both in the Coulomb blockade and the Aharonov-Bohm regimes \cite{PhysRevB.100.245130}. In this Letter, we demonstrate localization of quasiparticles with single-electron charge in the integer QH fillings of $\nu$ = 1, 2, and 6, as well as fractionally charged quasiparticles in $\nu = \pm 1/3$. To our knowledge, localization of the $\nu = - 1 /3$ state has not been previously reported. Due to the large energy scales in graphene, quantized charge and magnetic flux oscillations persist to temperatures up to two orders of magnitude higher than observed in GaAs-based 2DES previously studied. \par
\begin{figure*}
    \centering
    \includegraphics[width=\textwidth]{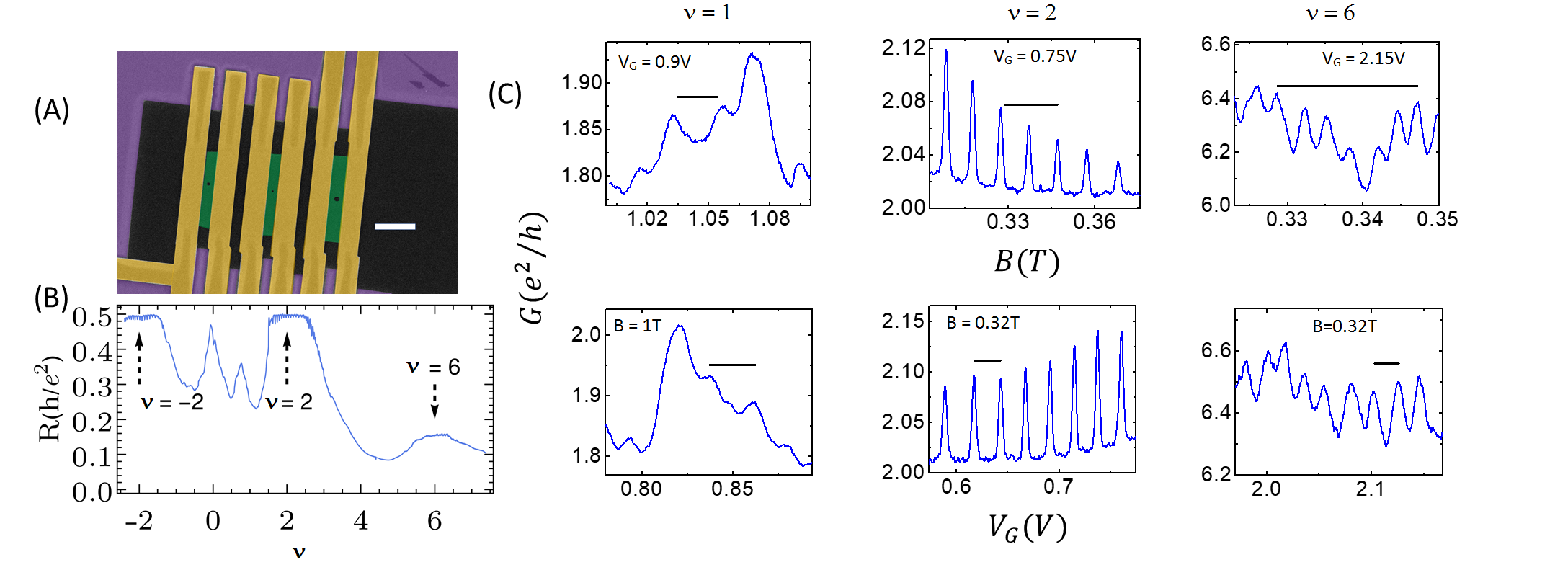}
    \caption{(A) False color SEM image of a typical sample with three devices fabricated on the same graphene flake. The white scale bar is 2 $\mu$m. (B) Integer QH resistance versus filling factor in Sample 1 at 320 mT and 700 mK. The fine, oscillatory features on the QH plateaus are from tunneling through the antidot. (C) Magnetic flux and charge oscillations in Sample 1, measured on (from left to right) $\nu = 1, 2, 6$ plateaus. The black scale bars on the back-gate dependence plots is 25 mV for all filling factors, corresponding to the addition of a single electron; while the scale bars on the magnetic field dependence plots is 20 mT corresponding to the addition of $\phi_0$. The effective radius calculated from the $\nu = 1$ magnetic field period is $r_{s1} \sim 250$ nm, in reasonable agreement with the design size.}
    \label{fig1}
\end{figure*}
Three high mobility ($\mu > 300,000$ cm$^2$V$^{-1}$s$^{-1}$) suspended graphene antidot samples were fabricated for this work (with examples shown in Fig.~\ref{fig1}A) using a procedure outlined in the SI \footnote{See supplementary information}. The diameters of the antidots studied, $d=100-300$ nm, were chosen to be significantly larger than the magnetic length, $l_B = \sqrt{\hbar/e B}$, yet small enough to form sizable quantum confined energy levels. The total channel length of graphene was chosen to be $L \sim 700$ nm, resulting an antidot-electrode distance of $\sim 200$ nm. Prior to the measurements, \textit{in situ} Joule heating was performed on all samples in order to minimize the contaminants on the graphene channels. After measurements, the samples were re-checked with a scanning electron microscope (SEM) to ensure that there was no structural damage to the sample from the Joule heating. We note that effective coupling between the electrodes and the antidot can be achieved in our devices despite the relatively large distance in between the electrodes and the antidot edge, as long as the edge modes are not too tightly confined to the physical edges. The coupling may also be enhanced due to the presence of a doping gradient near the electrodes, originating from the contact-induced potential profile in graphene \cite{PhysRevB.84.045414} or from impurities which cannot be effectively removed by Joule heating due to electrodes/substrate cooling. Sample 1 was fabricated to focus on the low field behavior of the integer filling factors, where the magnetic length is large and the QH plateaus are narrow. Samples 2 and 3 were fabricated with the aim to focus on the $\nu = 1/3$ state, which requires high magnetic field to be observed robustly (7 T in our devices) and has a much smaller QH energy gap compared to low integer filling factors at similar magnetic fields. Experimental details regarding the measurements are discussed in the SI. \par
At integer fillings, the addition of one flux quantum, $\phi_0=h/e$, into the antidot center corresponds to the expulsion of one electron per filled Landau level from the antidot edge. As we have shown previously for graphene \cite{PhysRevB.100.245130}, and has been demonstrated in other 2D systems, in the regime where Coulomb blockade physics is dominant this corresponds to $\nu$ conductance peaks per $\phi_0$, where $\nu$ is the filling factor of the graphene channel (Fig.~\ref{fig1}C). Similarly, the global back-gate can be used to attract or expel charge from the antidot edge, and the Coulomb interaction prevents the addition of more than one electron to the antidot edge at one time. In this way, the electron charge in the integer filling factors can be estimated using the average magnetic field period and back-gate voltage period of successive conductance peaks (see e.g. \cite{Goldman1010})
\begin{equation}
q = \phi_0 \frac{c \Delta V_G}{\Delta B_{\nu}},
\label{eq:charge}
\end{equation}
where $\Delta V_G$ is the back-gate voltage period, $\Delta B_{\nu}$ is the $\nu$ dependent magnetic field period, and $c$ is the gate capacitance per unit area which can be calculated from the positions of the QH plateaus with respect to back-gate voltage. For Sample 1 (Fig.~\ref{fig1}C), $c \sim 31.0$ aF$\mu$m$^{-2}$, and the back-gate voltage period $\Delta V_G \sim 24.7 $ mV and magnetic field period $\Delta B_{\nu=2} \sim 20$ mT (precisely measured for $\nu = 2$, but consistent for all other integer filling factors), give an estimate for the quasiparticle charge, $q \sim 0.99 \pm .07 e$, in good agreement with the expected charge for a single electron. This demonstrates the reliability of the QH antidot approach for measuring quasiparticle charge. We next apply the same method in the FQH regime for Samples 2 and 3. \par
\begin{figure}
    \centering
    \includegraphics[width=0.48\textwidth]{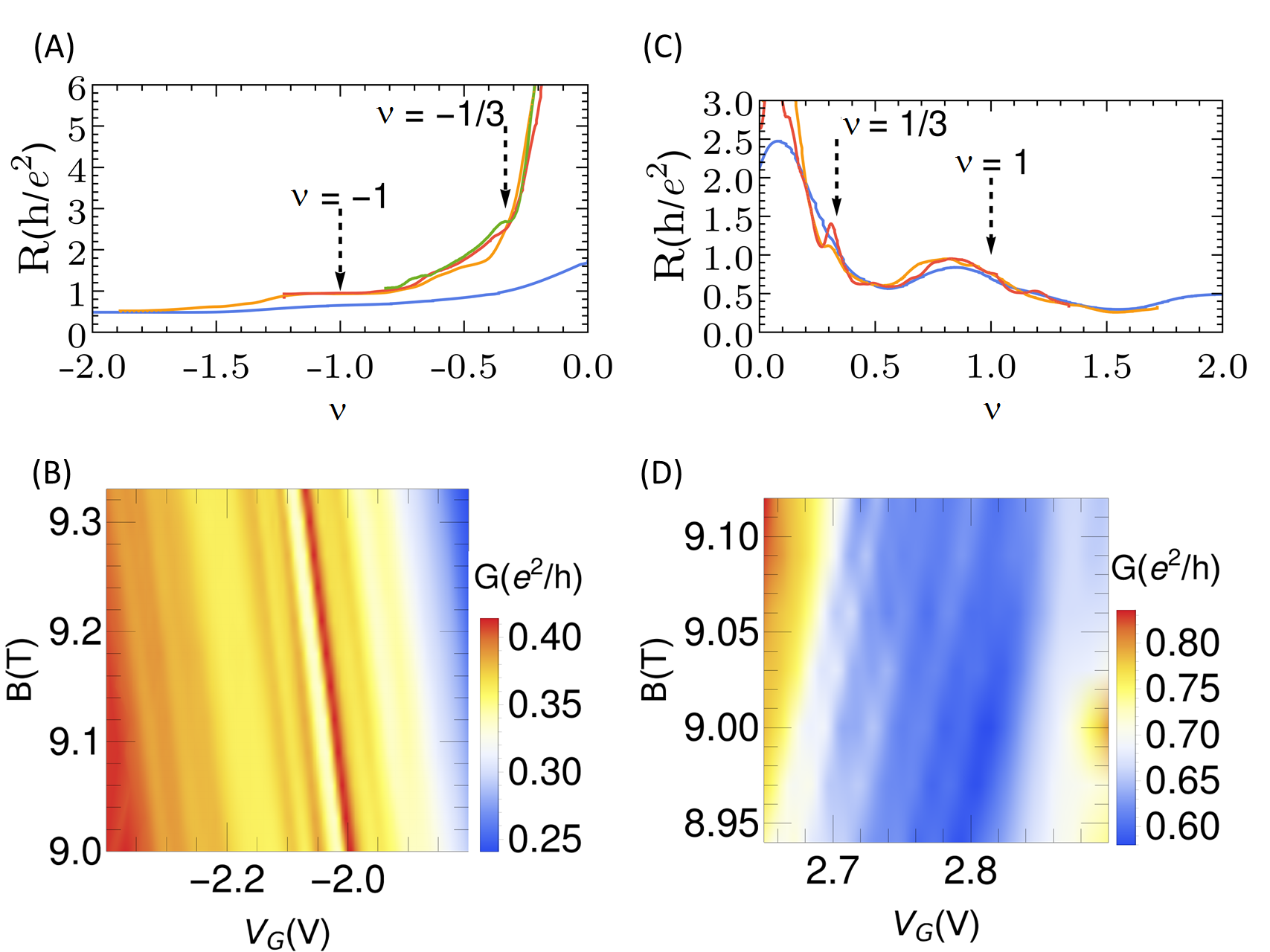}
    \caption{(A) Development of QH plateaus in Sample 2 with increasing magnetic fields: $B = $ 1, 4, 6, and 9 T. (B) Conductance vs. back-gate voltage and magnetic field for the $\nu = -1/3$ plateau for Sample 2 with an effective radius $r_{s2} \sim 75$ nm. (C) Development of QH plateaus in Sample 3 with increasing magnetic fields: $B = $ 3, 7, and 9 T. In this sample, the $\nu = 1/3$ plateau is not fully developed, and so the conductance is greater than $e^2 / 3 h$. (D) Conductance vs. back-gate voltage and magnetic field for the $\nu = 1/3$ plateau for Sample 3, which had an effective radius $r_{s3} \sim 110$ nm.
    \label{fig2} 
    }
\end{figure}
Edge modes encircling the antidot can be approximated by lowest Landau level (LLL) wavefunctions in the symmetric gauge, and the presence of the antidot confinement potential lifts the degeneracy of successive states \cite{PhysRevB.100.245130}. The $\nu=1/3$ state forms an incompressible electron liquid in the LLL with each electron state effectively occupied by the charge $e/3$, which also represents the charge of the quasiparticle excitations above the interacting ground state \cite{PhysRevLett.50.1395}. Therefore, in the $\nu=1/3$ regime, application of the back-gate voltage pushing single-electron states up or down in energy, adds one quasiparticle charge $q=e/3$ every period which corresponds to the shift of the energy spectrum by one electron state. Similarly, as a function of magnetic flux, the addition of $\phi_ 0$ to the antidot center shifts the electron spectrum by one state, leading to the expulsion of one quasiparticle from the antidot edge \cite{PhysRevLett.65.3369, PhysRevLett.66.806}, the same as in a QH interferometer \cite{PhysRevLett.98.106801}. \par
For Sample 2, the $\nu = -1/3$ plateau started to form at $\sim 4$ T and reached the quantized value of $e^2 / 3 h$ at 9 T (Fig.~\ref{fig2}A). On the plateau, conductance oscillations were observed with an average period of 54 mV in back-gate voltage that decayed rapidly when exiting the plateau. The sample also showed periodic behavior on  the $\nu = -1/3$ plateau with respect to magnetic field (Fig.~\ref{fig2}B) with a period of 240 mT. Using an area capacitance of $c \sim 62.0$ aF$\mu$m$^{-2}$ obtained from the back-gate dependence of the QH plateaus, it was found that $q = -0.36 \pm .09 e$ for the $\nu = -1 / 3$ quasiparticle charge. We note that the area capacitance in Sample 2 is significantly larger compared to Sample 1, due to the incomplete removal of silicon oxide underneath the graphene channel as well as channel sagging, resulting in a smaller graphene-gate distance. Separately calculating the antidot radius from the magnetic field period, $r_{\Delta B} = \sqrt{\phi_0/\pi \Delta B}$, and back-gate voltage period, $r_{\Delta V_G} = \sqrt{q/\pi c \Delta V_G}$, gives $r_{\Delta B} = r_{\Delta V_G} \equiv r_{s2} \sim 75$ nm, consistent with the physical size of the antidot measured using SEM imaging. Indeed, in high magnetic fields the effective size of the antidot is expected to be close to its physical size, as the magnetic length becomes much smaller than the antidot diameter. \par
Sample 3 showed the weak formation of a $\nu = 1/3$ plateau for magnetic fields above 7 T, but the two-terminal conductance did not reach $e^2 / 3 h$ at the fields measured (Fig.~\ref{fig2}C) due to limited sample quality. This plateau also showed periodic conductance oscillations in both back-gate voltage and magnetic field with periods of 34 mV and 95 mT respectively (Fig.~\ref{fig2}D). The area capacitance in this sample (without excess oxide and with much less channel sagging) was the same as Sample 1, $c \sim 31.0$ aF$\mu$m$^{-2}$, which gives $q = 0.29 \pm .05 e$ for the $\nu = 1 / 3$ quasiparticle charge, approximately the same magnitude as for $\nu = - 1 / 3$. The fractional quasiparticle charge measurements here have significantly larger uncertainties compared to previous studies on GaAs-based 2DEGs \cite{Goldman1010}, which may be attributed to the limited sample quality and higher measurement temperature in this work.\par
\begin{figure}
    \centering
    \includegraphics[width=0.48\textwidth]{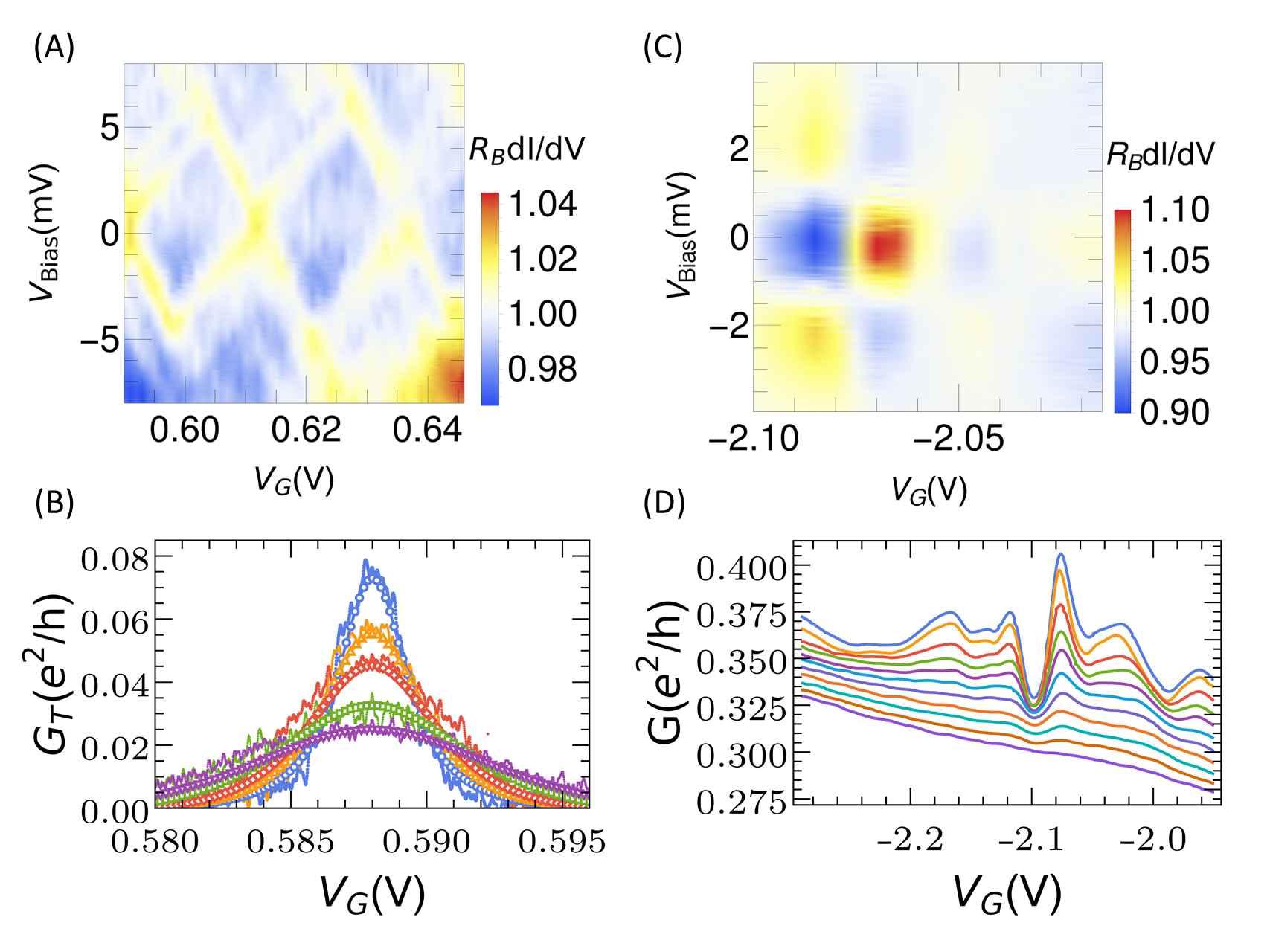}
    \caption{(A) Charge stability diagram of Sample 1 across three conductance peaks in the $\nu = 2$ plateau at 320 mT and 700 mK. Excited states are visible outside of the Coulomb gap. (B) Temperature dependence of a  conductance peak within $\nu = 2$ plateau, with best fit temperature progression overlaid, 0.7, 2.0, 3.0, 5.0, and 7.0 K. The tunneling conductance, $G_T$, is obtained by subtracting the bulk conductance (the QH plateau conductance) from the total two-terminal conductance. (C) Charge stability diagram of Sample 2 taken at 9.36 T, within the $\nu = -1/3$ plateau. Conductance is normalized by  $R_B$, a smooth bias-dependent background obtained by averaging over all differential conductance curves. (D) Conductance oscillations in $\nu = -1/3$ at increasing temperatures, from top to bottom 0.31, 0.53, 0.74, 1.05, 1.32, 1.52, 1.84, 2.10, 2.41, 2.68, and 2.90 K, taken at 9.36 T. Traces are successively offset $0.005$ $e^2/h$. \label{fig3}}
\end{figure}
Next, we investigate the energy scales and the temperature dependence of the conductance oscillations, first focusing on the structure of the $\nu = 2$ conductance peaks in Sample 1. The presence of Coulomb ``diamonds" in the charge stability diagram (Fig.~\ref{fig3}A) indicates that the dominant mechanism for charge transport is through tunneling, where the mutual Coulomb repulsion energy $U$ is the largest energy scale. The height of the Coulomb diamonds corresponds to $U$ or the addition energy $U + \Delta \varepsilon$ (where $U \sim 4$ meV and $\Delta \varepsilon \sim 1$ meV is the level spacing indicated by the lines outside the Coulomb diamonds), depending on the total spin of the antidot. In the following analysis, we neglect the effects of electron spin which is justified by the large ratio of $U/\Delta \varepsilon$, random fluctuations of the electrostatic potential as a function of back-gate voltage, and dephasing-induced broadening of the energy levels (see the discussion below). In principle, however, peak to peak separation is expected to alternate as a function of total electron spin and back-gate voltage. \par
It is evident that the bath temperature, $T \sim 700$ mK (60 $\mu$eV), cannot account for the broadening of the conductance peaks (FWHM $\sim 2.5$ mV in back-gate voltage as shown in (Fig.~\ref{fig3}B), or $\sim 500$ $\mu$eV in energy calculated from the ratio between back-gate voltage period and the corresponding Coulomb repulsion energy). While asymmetry in the tunneling probability between source and drain to the antidot can in principle lead to conductance peak broadening \cite{datta_1995}, such asymmetry needs to approach a factor of $\sim 50$ in order to account for the observed line-shape width and peak height and so is unlikely the broadening mechanism. To explain the conductance peak broadening, we adopt a low-frequency noise model in which fluctuations of the antidot confinement potential are caused by charges tunneling into/out of localized trapping sites. A similar model has been previously applied to charge qubits - see, e.g., \cite{RevModPhys.86.361, RevModPhys.60.537, Balandin2013}. For moving electrons, the low-frequency noise leads to a dephasing rate $\Gamma_{\phi} = K / 2 \hbar v$, where $K$ is the magnitude of pair-wise correlations of the noise potential, and $v$ is the edge mode velocity of the confined states \footnote{See supplementary information}. Based on the above observations, the parameters determining the tunneling behavior into the antidot  should satisfy $\Gamma_L, \Gamma_R, T < \Gamma_{\phi} < U, \Delta \varepsilon$; where $\Gamma_{L,R}$ are the tunneling rates for source and drain, and $U + \Delta \varepsilon \sim 5$ meV. In this regime, the resonant peak follows a Breit-Wigner formula \cite{PhysRevB.44.6199, PhysRevB.44.1646}. Furthermore, the highest temperature, $T \sim 7$ K (600 $\mu$eV), does not exceed the level spacing or charging energy. So approximately only one non-degenerate level participates in charge transport through the antidot at the temperatures studied here. The tunneling conductance is then given by
\begin{equation}
G(\mu) = \frac{e^2}{\pi \hbar} \frac{\Gamma}{4 k_B T}\frac{\Gamma_L \Gamma_R}{\Gamma_L+ \Gamma_R} \int d\varepsilon \frac{\cosh^{-2}(\mu - \varepsilon/ 2 k_B T)}{\varepsilon^2 + \Gamma^2},
\label{tdep_eq}
\end{equation}
where $\Gamma = (1/2)(\Gamma_L + \Gamma_R) + \Gamma_{\phi}$. Assuming symmetric tunneling rates for source and drain, the conductance peak at the lowest temperature, $T \sim 700$ mK (60 $\mu$eV), is best fit with tunneling rates $\Gamma_L =  \Gamma_R \sim 25$ $\mu$eV and an upper-bound dephasing rate $\Gamma_{\phi} \sim 291$ $\mu$eV. Using these parameters, conductance peaks at higher temperatures calculated from~\eqref{tdep_eq} are in good agreement with the measurements. We note that the edge mode velocity, $v = \Delta \varepsilon r_{s1}/\hbar \sim 3 \times 10^5$ m/s, is among the largest values achieved in 2DES, which is a consequence of both the sharpness of the confinement potential and the large Fermi velocity of graphene.\par
We further discuss the bias and temperature dependence of the $\nu=-1/3$ conductance oscillations in Sample 2 in order to estimate the confinement energy scales of the fractional quasiparticles. The conductance traces as a function of back-gate voltage were significantly broadened due to a combination of dephasing and strong tunneling, while the charge stability diagram (Fig.~\ref{fig3}C) shows a checkerboard pattern similar to the one known in the case of coherent electron transport in a Fabry-Perot interferometer (FPI), see, e.g., \cite{PhysRevLett.103.206806,Nakamura2019}. Although the geometry of our antidot structure is different from an FPI, for strong antidot-electrode tunneling the transport pattern in the antidot can be viewed as interference of the two tunneling points. The observation of the checkerboard pattern implies then that the quasiparticle transport through our antidot is coherent, $\Delta \varepsilon > \Gamma_{\phi}$, and is not dominated by Coulomb repulsion, as in, e.g., (Fig.~\ref{fig3}A). This conclusion can be supported by the fact that the quasiparticle interaction energy is a factor of 9 smaller than the electron repulsion energy $U$, and is further screened by the contacts. \par
We also note that the stripes shown in (Fig.~\ref{fig2}D) for the electron branch in an antidot sample have positive slope regardless of the strength of the Coulomb repulsion. In contrast, the slope in the non-interacting regime would be negative in the case of an FPI -- see \cite{PhysRevB.79.241304,Nakamura2019}. This difference stems from a simple geometric difference in the 2D confining potential, the cross-sectional area of which increases with energy in the case of an FPI, and decreases with energy in the case of an antidot. Of course, the slope changes sign with the changing sign of the charge, $q$, of tunneling particles, because of the changing sign of the energy shift $qV_G$ induced by the back-gate. This is demonstrated in (Fig.~\ref{fig2}B) and (Fig.~\ref{fig2}D) for "fractional hole" and "fractional electron", respectively.\par
Traces taken at successively higher temperatures show the conductance oscillations to be resolvable up to temperatures in excess of 2 K (Fig.~\ref{fig3}D). The checkerboard spacing of $\Delta V_{\text{Bias}} \sim 2$ mV can be used to estimate the edge mode velocity: $v = q \Delta V_{\text{Bias}} r_{s2} / \hbar$. Since the precise structure of the tunneling contacts is not known, the magnitude of the charge q coupling to the bias voltage is uncertain. Using a minimum value of $q = e/3$, we obtain a lower-bound edge mode velocity of $\sim 7 \times 10^4$ m/s, again among the highest values achieved in similar devices.\par
In conclusion, we have demonstrated localization of QH quasiparticles in suspended graphene antidots samples, in integer fillings $\nu$ = 1, 2, and 6, as well as fractional fillings $\nu = \pm 1 / 3$.  In both regimes, graphene showed robust single charge and flux oscillations with consistently larger energy scales than previously achieved in GaAs-based 2DES. Our work establishes a basis for further investigations, including utilizing high quality graphene-based QH antidots (e.g., using suspended or hexagonal boron nitride-encapsulated graphene)  to study braiding of fractionally charged anyons \cite{PhysRevLett.99.096801}, or to fabricate novel quantum information processing devices.\par
S.M.M. would like to acknowledge helpful discussions with V.J. Goldman and Y.-M. Zhong, as well as assistance with sample preparation from X. Chen and J. Zhang. The authors acknowledge support from the N.S.F. under award DMR-1836707.\par

\end{document}